# Novel Synthesis and High Pressure Behavior of $Na_{0.3}CoO_2$ x 1.3 $H_2O$ and Related Phases.

Sangmoon Park[1], Yongjae Lee[1], Arnie Moodenbaugh[2] & Thomas Vogt [1,3,*]
[1]Physics Department, Brookhaven National Laboratory, Upton, NY 11973-5000
[2]Materials Science Department., Brookhaven National Laboratory, Upton, NY 11973-5000
[3]Center for Functional Nanomaterials, Brookhaven National Laboratory, Upton, NY 11973-5000

***We have prepared powder samples of $Na_xCoO_2$ x $yH_2O$ using a new synthesis route. Superconductivity was observed in $Na_{0.3}CoO_2$ x $1.3H_2O$ between 4 and 5K as indicated by the magnetic susceptibility. The bulk compressibilities of $Na_{0.3}CoO_2$ x $1.3H_2O$, $Na_{0.3}CoO_2$ x $0.6H_2O$ and $Na_{0.3}CoO_2$ were determined using a diamond anvil cell and synchrotron powder diffraction. Chemical changes occurring under pressure when using different pressure transmitting media are discussed and further transport measurements are advocated.***

The observation of superconductivity in $Na_{0.3}CoO_2$ x $1.3H_2O$ below 5K [1] has sparked interest in this system since this is the second known case where superconductivity arises from doping a Mott insulator ($CoO_2$). $CoO_2$ is a frustrated triangular spin system and provided the initial line of reasoning for Anderson's RVB model [2]. A number of authors propose a d-wave-type pairing based on these ideas [3], while others argue that the proximity to a ferromagnetic state favor a p-wave-pairing mechanism [4]. The crystal structures of the $Na_xCoO_2$ x$yH_2O$ family are built up of hexagonal layers composed of x nonmagnetic $Co^{3+}$ and 1-x low-spin $Co^{4+}$ (S=1/2) ions which are separated by a "charge reservoir" of $Na_x(H_2O)_y$. The role of this "charge reservoir" is not at all clear and further chemical and physical modifications are required to probe the stability field of this superconducting family. We show here that pressure-induced intercalation can be used to alter and distort the hexagonal structure thus providing us an opportunity to probe the structural prerequisites for super-conductivity in this family of compounds.

Early on it was realized that these materials are extremely sensitive to variations of their water content when handled without special precautions in the atmosphere [5]. Furthermore, the original synthesis of $Na_{0.3}CoO_2$ x $1.3H_2O$ [1] relies on the de-intercalation of Na using Br ions in acetonitrile (flash point = 42$^o$F). In attempts to scale up the amounts of product and avoid the associated environmental hazards which are associated when manipulating large amounts of high concentrations of bromine, new synthetic routes are called for. Chou et al [6] used an electrochemical route for the de-intercalation step which allows a better control of the Na content. We report here that superconducting powder samples of large quantities can be made by using $Na_2S_2O_8$ for the oxidation and de-intercalation step: the precursor material $Na_{0.7}CoO_2$ was obtained by heating a mixture of $Na_2CO_3$ with a 10 mol% excess (Alfa & Fischer 99.5%) and $Co_3O_4$ (Alfa 99.7%) at 850$^o$C for 8 h under $O_2$ (g) flow. $Na_xCoO_2$ x $yH_2O$ was then made using $Na_2S_2O_8$ in aqueous solution using an equimolar ratio with $Na_{0.7}CoO_2$ by stirring for 22h in a beaker covered with a Parafilm$^{TM}$. For more mechanistic details see [7,8]. Using 4-5 drops of 1N $NH_4OH$ in 20ml DI water (pH~10.5) allowed us to reproducibly obtain $Na_{0.3}CoO_2$ x $1.3H_2O$. Figure 1 shows that the variation of pH is crucial for obtaining the superconducting phase. An optimal pH to obtain the superconducting $Na_{0.3}CoO_2$ x $1.3H_2O$ phase was found to be close to 10.5. The advantages of this synthesis are that large amounts of sample can be made; it is environmentally benign and uses only water as a solvent. All samples were placed in humidified containers and characterized by X-ray powder diffraction. For comparison we also prepared samples using bromine for the de-intercalation/oxidation step. A SQUID magnetometer (Quantum Design) was used to determine the magnetic susceptibility as a function of temperature. Figure 2 compares a sample of $Na_{0.3}CoO_2$ x $1.3H_2O$ made via the $Na_2S_2O_8$ route (a) with one made using the bromine/acetonitrile route (b). We observe a slightly earlier onset of superconductivity in the latter and attribute this to minute variations of the Na content [7].

All samples used for the high-pressure x-ray powder diffraction studies were made using the $Na_2S_2O_8$ synthesis route. In-situ high pressure powder diffraction experiments were performed using a diamond anvil cell (DAC) at beamline X7A at the National Synchrotron Light Source (NSLS) at Brookhaven National Laboratory. The detailed setup is described elsewhere [8]. Due to the extreme moisture sensitivity of $Na_{0.3}CoO_2$ x $1.3H_2O$ it was necessary to contain the sample in a wet environment prior to loading in the diamond anvil cell. Furthermore, care was taken to minimize the exposure time to the atmosphere during loading. Initially a methanol:ethanol:water mixture of 16:3:1 by volume was used as a pressure transmitting fluid to ensure hydrostaticity. However, we noticed a phase transition from the hexagonal to a monoclinic phase already at very low pressures (~0.15GPa) (Figure 3) as well as an initial increase of the a axis of the monoclinic cell. When changing the pressure transmitting fluid to Fluorinert$^{TM}$ no phase transition or intercalation under pressure is observed. It is noteworthy that this intercalation affects the basal plane of the monoclinic distorted hexagonal unit cell and not the c-axis. This could be related to an increase of the Na coordination number from six to seven within the charge reservoir as is frequently observed in Na containing zeolites under conditions where pressure-induced hydration occurs [9].The pressure the sample was subjected to was measured by detecting the shift of the R1 emission line of the included ruby chips [10]. In our experiments, no evidence of nonhydrostatic behaviour or pressure anisotropy was detected, and the R1 peaks of 3 to 4 included ruby chips remained strong and sharp with deviations less than ± 0.1GPa. The hydrostatic limit for Fluorinert$^{TM}$ is generally quoted to be ca. 1.5 GPa [11,12], but others report extended hydrostaticity when the samples are softer than the glass that Fluorinert forms under pressure [13]. Flourinert $^{TM}$ was also used in the determination of the effect of hydrostatic pressure on the superconducting transition temperature dlnTc/dp by Lorenz et al [14]. Bulk moduli were determined by fitting the normalized volumes to a second-order Birch-Murnaghan Equation of State [15] using a fixed pressure derivative of 4.

The derived bulk compressibilities of 43(2) GPa, 90(6)GPa and 101(3)GPa for {x=0.3,y=1.3}, {x=0.3,y=0.6} and{x=0.3,y=0} respectively show the expected higher compressibility of the superconductor compared to the doped metal oxide {x=0.3,y=0} as well as the intermediate {x=0.3,y=0.6} oxyhydrate (Figure 4). Lorenz et al [14] showed that dTc/dp is negative and nonlinear up to 1.6GPa. Interestingly enough dlnTc/dp ~- 0.07GPa is as was pointed out similar to values observed in electron-doped cuprates [14]. Despite the low Tc $Na_{0.3}CoO_2$ x $1.3H_2O$ appears to behave according to the universal relationship between dlnTc/dp and Tc [16] established for high-Tc superconductors. The structures show a very strong anisotropy under pressure: in all cases the c-axis is the most compressible direction, whereas the a-axes show only a marginal decrease with pressure (see figure 4 for details). However, when intercalating extra molecules under pressure (alcohols or water), the structure distorts and the former hexagonal basal plane expands slightly under pressure in both $Na_{0.3}CoO_2$ x $1.3H_2O$ and $Na_{0.3}CoO_2$ x $0.6H_2O$. The compressibility of the monoclinic phase measured when using the methanol:ethanol:water mixture as pressure transmitting fluid is about 30% higher $K_0$=60(3)GPa) than the one obtained for the hexagonal phase using Flourinert ™ ($K_0$=43(2)GPa). This is consistent with other observed phases altered by intercalation under pressure [9,13]. The individual unit cell compressibilities indicate that the monoclinic phase is much less compressible within the $CoO_2$ layers having values quite similar to $Na_{0.3}CoO_2$. We encourage to repeat the measurements of the pressure dependence on Tc using methanol:ethanol:water as pressure transmitting medium to probe if the observed monoclinic phase is also superconducting and understand how the expansion of the a,b plane influences the electronic properties.

In summary we have found a new environmentally benign synthesis route for the superconductors of the $Na_xCoO_2$ $yH_2O$ family (x~0.3). Furthermore, we have determined the intrinsic bulk compressibilities of $Na_{0.3}CoO_2$ x $1.3H_2O$, $Na_{0.3}CoO_2$ x $0.6H_2O$ and $Na_{0.3}CoO_2$. A monoclinic distortion of $Na_{0.3}CoO_2$ x $1.3H_2O$ was found to occur at very low pressures when using an alcohol:water mixture as a pressure transmitting fluid. Magnetic susceptibility and resistivity measurements of this pressure-stabilized phase could provide us valuable information about this fascinating new family of superconductors and its stability field.

This work was supported by an LDRD from BNL. The authors thank J. Hu and the Geophysical Laboratory for the access to their ruby laser system at beamline X17C. Research carried out in part at the NSLS at BNL is supported by the U.S. DOE (DE-Ac02-98CH10886 for beamline X7A).

Figure captions :

**Figure 1:**
Powder X-ray diffraction patterns (Cu Kα radiation) for $Na_xCoO_2$ $yH_2O$ phases prepared using $Na_2S_2O_8$ as a function of pH.

**Figure 2:**
Magnetic susceptibility of $Na_{0.3}CoO_2$ $1.3H_2O$ prepared via the $Na_2S_2O_8$ synthesis route (a) and using $Br_2$/acteonitrile (b).

**Figure 3:**
Pressure-induced changes of the x-ray powder diffraction patterns of $Na_{0.3}CoO_2$ x $1.3H_2O$ when using an alcohol/water mixture as a pressure-transmitting fluid.

**Figure 4:**
Unit cell volume and axis compressibilities of (a) $Na_{0.3}CoO_2$ x $1.3H_2O$, (b) $Na_{0.3}CoO_2$ x $0.6H_2O$ and (c) $Na_{0.3}CoO_2$. The hydrates were measured using Flourinert TM(data in red) and an alcohol/water mixture as pressure transmitting fluid (data in black).

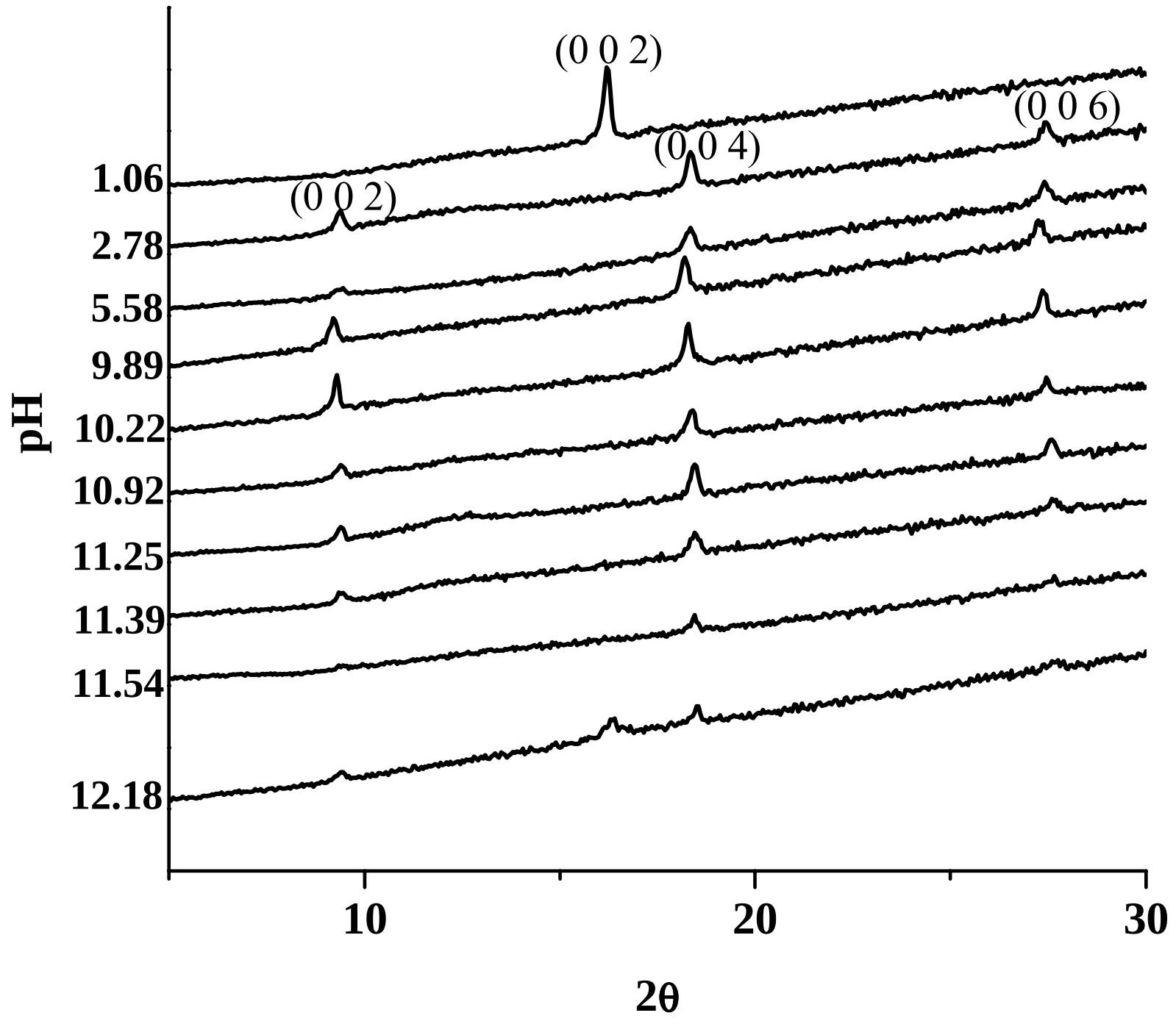

Fig. 1

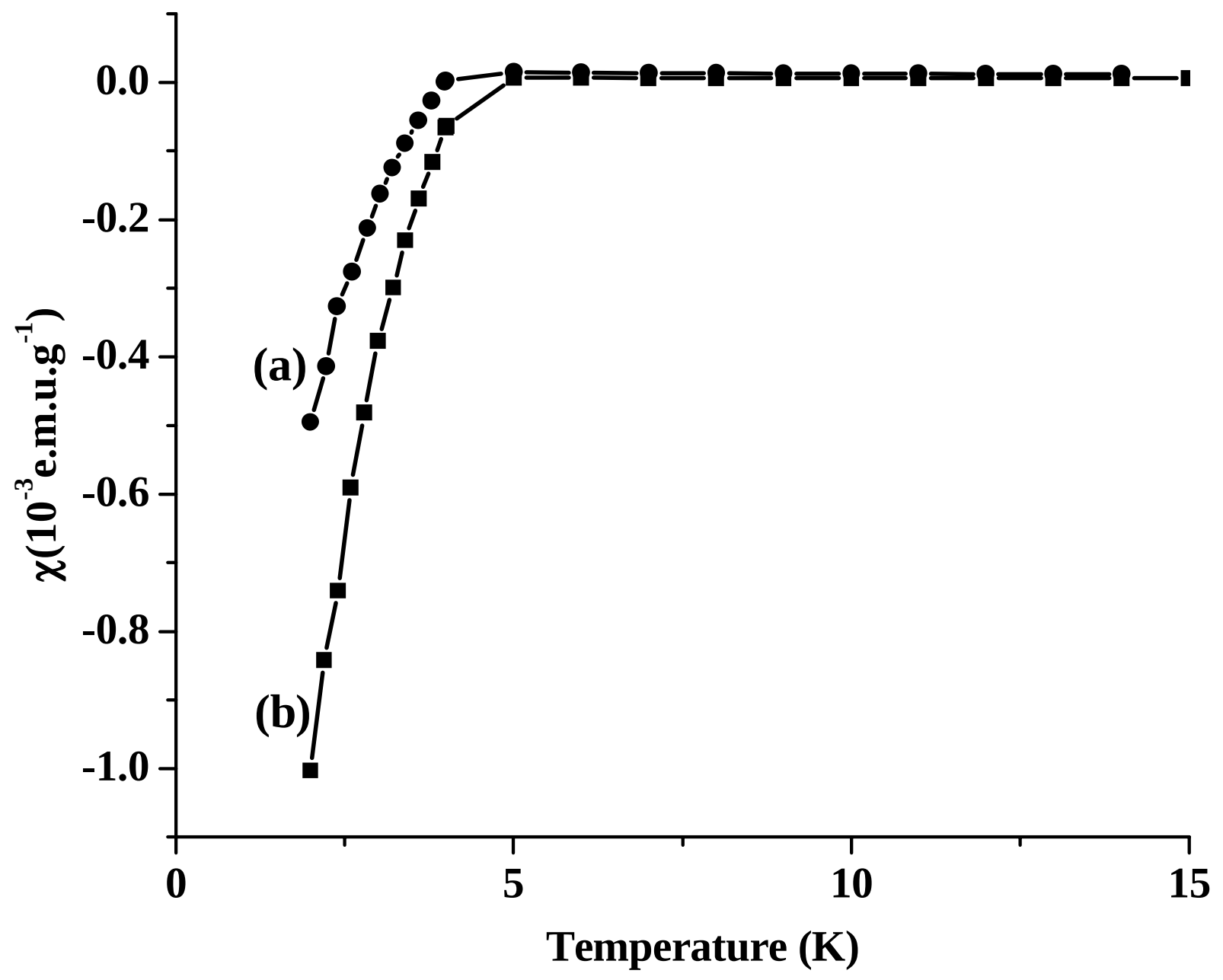

Fig. 2

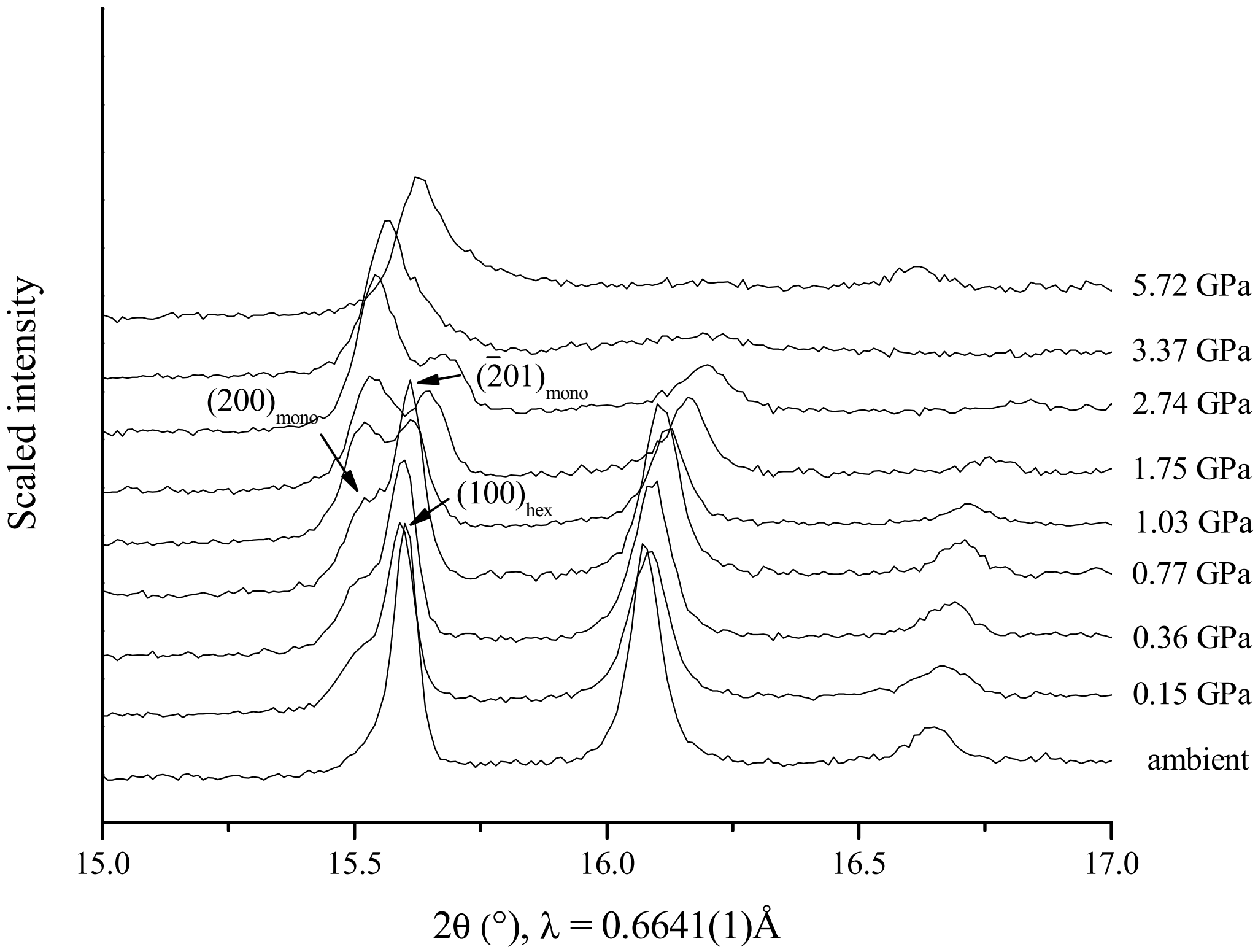

Fig. 3

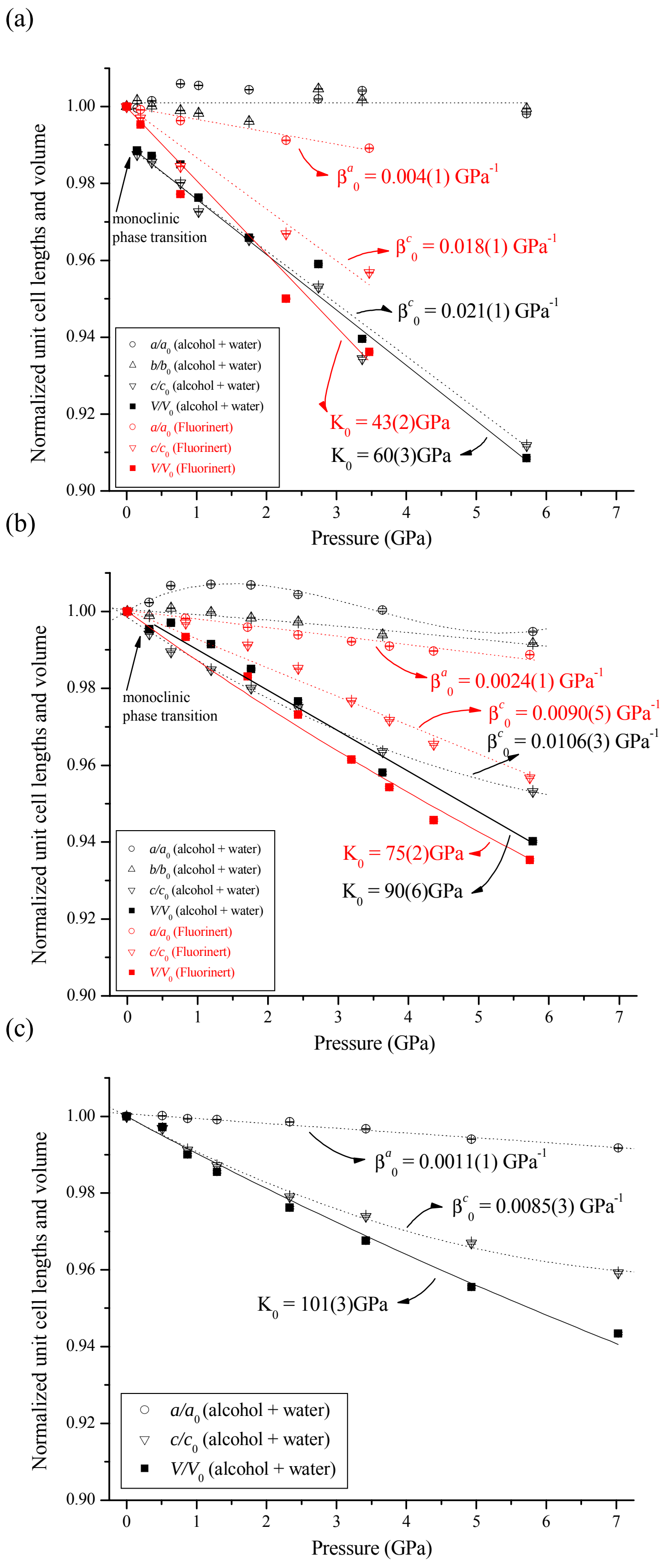

(a)
Normalized unit cell lengths and volume
Pressure (GPa)
β^a_0 = 0.004(1) GPa^-1
β^c_0 = 0.018(1) GPa^-1
β^c_0 = 0.021(1) GPa^-1
K_0 = 43(2)GPa
K_0 = 60(3)GPa
monoclinic
phase transition
a/a_0 (alcohol + water)
b/b_0 (alcohol + water)
c/c_0 (alcohol + water)
V/V_0 (alcohol + water)
a/a_0 (Fluorinert)
c/c_0 (Fluorinert)
V/V_0 (Fluorinert)
(b)
Normalized unit cell lengths and volume
Pressure (GPa)
β^a_0 = 0.0024(1) GPa^-1
β^c_0 = 0.0090(5) GPa^-1
β^c_0 = 0.0106(3) GPa^-1
K_0 = 75(2)GPa
K_0 = 90(6)GPa
monoclinic
phase transition
a/a_0 (alcohol + water)
b/b_0 (alcohol + water)
c/c_0 (alcohol + water)
V/V_0 (alcohol + water)
a/a_0 (Fluorinert)
c/c_0 (Fluorinert)
V/V_0 (Fluorinert)
(c)
Normalized unit cell lengths and volume
Pressure (GPa)
β^a_0 = 0.0011(1) GPa^-1
β^c_0 = 0.0085(3) GPa^-1
K_0 = 101(3)GPa
a/a_0 (alcohol + water)
c/c_0 (alcohol + water)
V/V_0 (alcohol + water)

Fig. 4